\documentclass[letterpaper, 11pt]{article}
\usepackage{latexsym}
\usepackage{amssymb}
\usepackage{times}
\usepackage[in]{fullpage}
\usepackage{amsmath,amsfonts,amsthm}

\newcommand{\ble}{\begin{lemma}}
\newcommand{\ele}{\end{lemma}}

\newtheorem{lemma}{Lemma}[section]

\newtheorem{theorem}[lemma]{Theorem}

\newtheorem{corollary}[lemma]{Corollary}

\newcommand{\beao}{\begin{eqnarray*}}
\newcommand{\eeao}{\end{eqnarray*}\noindent}
\newcommand{\beam}{\begin{eqnarray}}
\newcommand{\eeam}{\end{eqnarray}\noindent}

\begin{document}

\title{Succinct Sampling on Streams}
\author{Vladimir Braverman, Rafail Ostrovsky, Carlo Zaniolo\\
  \textrm{University of California Los Angeles}\\
  \texttt{\{vova, rafail, zaniolo\}@cs.ucla.edu}}

\maketitle

\thispagestyle{empty}

\begin{abstract}
A streaming model is one where data items arrive over long period of
time, either one item at a time or in bursts.  Typical tasks include
computing various statistics over a sliding window of some fixed
time horizon. What makes the streaming model interesting is that as
the time progresses, old items expire and new ones arrive.  One of
the simplest and most central tasks in this model is sampling. That is,
the task of maintaining up to $k$ uniformly distributed items from a
current time-window. We call
sampling algorithms {\bf succinct} if they use provably optimal (up
to constant factors) {\bf worst-case} memory to maintain $k$ items
(either with or without replacement). We stress that in many
applications, structures that have {\em expected} succinct
representation as the time progresses are not sufficient. That is,
expected small memory solutions in a streaming environment will
never provide a fixed bounded memory guarantee over the lifetime of a
(very large) stream, as small probability events eventually happen
with probability 1. Thus, in this paper we ask the following
question: Are succinct sampling on streams (or $S^3$ algorithms)
possible, and if so, for what models?

Perhaps somewhat surprisingly, we show that $S^3$ algorithms
(i.e. with matching upper and lower bounds and worst case fixed memory guarantees) are
possible for {\em all} variants of the problem mentioned above, i.e.,
both with and without replacement and both for one-at-a-time and
bursty arrival models. In addition to fixed memory guarantees, our
solution has additional benefits that are important in
applications: in ``one-item-at-a-time'' model, the samples produced
over non-overlapping windows are completely independent of each
other (this was not the case for previous solutions) and in the
bursty model, previous solutions required floating point
computations; we do not. Finally, we use $S^3$ algorithms to solve
various problems in the sliding windows model, including frequency
moments, counting triangles, entropy and density estimations. For
these problems we present \emph{first} solutions with provable
worst-case memory guarantees. The results that we arrive at are based
on the novel sampling method that could be of independent interest.
\end{abstract}

\newpage
\setcounter{page}{1}

\section{Introduction}

A \emph{data stream}  is an ordered, possibly infinite, set of
elements, $p_1, p_2, \dots, p_N, \dots$, that can be observed only
once. The data stream model recently became extremely useful for numerous
applications including networking, finance, security,
telecommunications, world wide web, and sensor monitoring. Since
data streams are unbounded, it is impossible to store all data and
analyze it off-line using multiple passes (in contrast to
traditional database systems). As a result, precise calculation of
some queries or statistics may be infeasible, and approximate
solutions are provided. One of the main challenges is to minimize
memory requirements, while keeping a desirably precise answer.

Many applications are interested in analyzing only recent data
instead of all previously seen elements. The sliding window model,
introduced by Babcock, Babu, Datar, Motwani and Widom
\cite{models_issues}, reflects this interest. In this model we
separate past elements into two sets. The most recent elements represent
a window of \emph{active} elements, whereas others are
\emph{expired}. An active element may eventually become expired, but
expired elements stay in this status forever. The sliding window is a
set of all currently active elements, i.e., $W = \{p_{N-n},\dots,
p_N\}$, where $N$ is the current size of a stream and $n$ is the
number of active elements frequently refereed to as window's size. Only
active elements are relevant for statistics or queries. For
\emph{the sequence-based} model, the window size is predefined and does
not depend on the current status of the stream. For
\emph{the timestamp-based} model, each element $p$ is associated with
a non-decreasing \emph{timestamp}, $T(p)$. An element is active if
$t-T(p) < t_0$, where $t$ is the current timestamp and $t_0$ is some
predefined and fixed value. Thus, window's size strictly depends on
$t$ and can be any non-negative number. We refer readers to the
works of Babcock, Babu, Datar, Motwani and Widom
\cite{models_issues}, Muthukrishnan \cite{strbook} and Aggarwal
\cite{strbook1} for more detailed discussions of these models, and
related problems and algorithms.

\subsection{Questions posed and our results}
Random sampling methods are widely used in data stream processing,
because of their simplicity and efficiency. What makes these methods
attractive for many applications is that they store elements instead
of synopses, allowing us to change queries in an ad hoc manner and reuse
samples a posteriori, with different algorithms. Further, sampling
methods are natural for streams with multi-dimensional elements,
while other methods, such as sketches, wavelets and histograms, are
not easily extended to multi-dimensional cases. We refer readers to
the recent surveys by Datar and Motwani \cite{strbook1} (Chapter 9) and
Muthukrishnan \cite{strbook} for deeper discussions of these and
other advantages of sampling methods. What makes sampling
non-trivial is that the domain's size changes constantly, as well as the
probabilities associated with elements. We distinguish between
sampling \emph{with replacement} where all samples are independent
(and thus can be repeated), and its generalization, sampling
\emph{without replacement}, where repetitions are prohibited. Due to
its fundamental nature, the problem has received considerable attention
in the last decades. Vitter \cite{reservoir} presented reservoir
sampling, probably the first algorithm for uniform sampling (with
and without replacement) over streams.
A reservoir is an array with size $k$ where the current samples are
stored. We choose $p_i$ to be a sample w.p. $1$, if $0\le i<k$ and
w.p. $k\over {i+1}$ otherwise. If $p_i$ is chosen and there is no
space in the reservoir, we delete one of the previously chosen
samples and put $p_i$ instead. This algorithm requires $\Theta(k)$
memory and generates uniform random sample without replacement of
size $k$. Numerous sampling methods were
developed for different scenarios and distributions. These works
include, among many others, concise and counting samplings by
Gibbons and Matias \cite{mg}; priority sampling by Duffield, Lund and
Thorup  \cite{priority1}, Alon, Duffield, Lund and Thorup
\cite{priority2}, and Szegedy \cite{dlt}; weighted sampling by
Chaudhuri, Motwani and Narasayya \cite{joins}; faster reservoir
sampling by Li \cite{reservoir1}; density sampling by Palmer and
Faloutsos \cite{density_sampling}; and non-uniform reservoir sampling
by Kolonko and W\"{a}sch \cite{reservoir2}. Several data stream
models (including sliding windows) allow deletions of stale data. In
these models sampling becomes even more challenging, since samples
eventually expire and must be replaced. The recent works include
chain and priority samplings by Babcock, Datar and Motwani
\cite{sampling}; biased sampling by Aggarwal \cite{biased};
Aggarwal, Han, Wang, and Yu \cite{more-biased}; sampling in dynamic
streams by Frahling, Indyk, Sohler \cite{dynamic}; and inverse
sampling by Cormode, Muthukrishnan and Rozenbaum \cite{inverse}.

\subsubsection{Historic perspective on sampling on sliding windows}
%%%%%%%%%%%%% NEW %%%%%%%%%%%%%%%%%%%%%%%%%%%%%%%%5
In this paper we address the problem of maintaining a random sample of fixed size $k$ for every window.
This problem was introduced in the pioneering paper of Babcock, Datar and Motwani
\cite{sampling} and several solutions are known.
One possible and well-known method, described in this paper, is periodic or systematic sampling.
A sample $p_i$ is picked from the first $n$ elements, defining the sequence of all its replacements as
$p_{i+sn}, s=1,2,\dots$. This method provides a deterministic solution
for the sampling problem and uses $O(k)$ memory.
However, it was criticized for its inability to deal with periodic data and vulnerability to malicious behavior.
For more detailed criticism of periodic sampling see, for example, the papers of
Duffield \cite{sampling_networks} and Paxson, Almes, Mahdavi and Mathis \cite{networks}.
Therefore, the following requirement is important: samples from distinct
windows should have either weak or no dependency.

Babcock, Datar and Motwani
\cite{sampling} provided the first
effective algorithms that do not possess periodic behavior.
The chain algorithm provides samples from sequence-based windows.
The algorithm picks every new element w.p.
$1\over n$. In addition, for every chosen element, it uniformly
selects and stores its replacement from its $n$ successors, creating
a chain of replacements. If a sample expires, its successor in the
chain becomes the new sample. The expected size of this chain is
constant and $O(\log(n))$ with probability $1-{1\over n^c}$ for a
constant $c$. Repeating this $k$ times gives sampling an expected
optimal memory $O(k)$ and a high-probability upper bound
$O(k\log{n})$. Priority sampling provides samples from timestamp-based windows.
The algorithm associates each element with a
priority, a real number randomly chosen from $[0,1]$. An element is
chosen as a sample if its priority is highest among all active
elements. The algorithm stores only elements that may potentially
become samples. The expected number of such elements is $O(\log{n})$
and is the same with probability $1-{1\over n^c}$ for a constant
$c$. Therefore, the expected and high-probability memory is
$O(k\log{n})$. The authors also note that sampling without replacement
can be simulated, with high probability, by over-sampling with replacement.
%
%
%in terms of the number of non-expired elements and it stays in this
%range with high probability. Our $S^3$-algorithms improve the
%results of Babcock, Datar and Motwani, providing worst-case memory
%guarantees.

%Surprisingly, no solution exists (i.e. one that  satisfies all
%requirements.) Such succinct algorithms as reservoir sampling
%\cite{reservoir} by Vitter do not address sliding windows.
%Algorithms presented by Babcock, Datar and Motwani in
%\cite{sampling} address sliding window, but they are not succinct
%(i.e, their memory requirements will grow with the size of the
%stream, as small probability events will eventually require more and
%more memory). A naive solution for sequence-based windows is to pick
%samples randomly from the first window and replace expired element
%with the new one. This solution does not satisfy the independence
%requirement.

Recently, Zhang, Li, Yu, Wang and Jiang \cite{sampling1} provided sampling algorithms for sliding windows.
These algorithms use linear memory, work only for small windows,
and therefore cannot be compared to our results.

\subsubsection{Our contribution}
In this paper we optimally solve the problem of sampling on sliding windows.
We refine the requirements above, defining two critical properties
of sampling algorithms for sliding windows. First, they must use
provably optimal memory. Second, we require complete independence for non-overlapping windows,
refining the ideas from \cite{sampling}. By \emph{Succinct
Sampling on Streams (or $S^3$)} we denote an algorithm that satisfies the
above requirements. In this paper we ask the following question:
\emph{Are $S^3$ algorithms possible, and if so, for what models}? We
show that $S^3$ algorithms for uniform distribution are possible for
all variants of the problem, i.e., both with and without replacement
and both for sequence and timestamp-based windows. That is, for all
these models we present a matching upper and lower bounds. For
sequence-based windows, we use $\Theta(k)$ memory to generate a
sample with size $k$, both with and without replacement. For
timestamp-based windows, we present $\Theta(k\log{n})$ algorithms,
where $n$ is the number of non-expired elements, and prove that this
is an optimal solution.

    From the theoretical perspective, $S^3$ algorithms are important since
their improvements over the chain and priority methods can be arbitrarily large, in the worst case.
That is, for any $m=o(n)$ there exists a data stream $D$ such that the ratio between
maximal memory over all windows for the previous solutions and $S^3$ algorithms is at least $m$.
Consider, for instance, chain sampling where a chain of replacement is maintained.
The probability that each of $k$ chains in \cite{sampling} has length $m$ (or larger) is at least $1/m^{km}$.
Therefore, for streams with size $N \ge m^{mk}$, the expected maximal memory usage over all windows
is at least $\Omega(mk)$. In contrast, our method requires $O(k)$ memory.
A similar result can be obtained for timestamp-based windows.

One may argue that in practice the requirement of producing a sample for every window is too rigid.
The relaxed version of the problem, where samples are outputted for all windows except a small fraction, is
sufficient. Indeed, any statistic that is based on sampling accepts an error with small probability.
Thus, by cutting off the windows with large chains, we can maintains the desired statistics
and only increase the probability of error by a negligible amount.
Moreover, if the size of the entire stream is polynomial in the window size, i.e., $N = poly(n)$,
then the probability of error on the whole stream can be made as small as $1/poly(N)$.
As a result, one may claim that the improvement provided in our paper is incremental,
from the practical perspective. This argument is flawed for two important reasons.

First, this argument misses the important point that for a constant number
of samples, our algorithms are asymptotically superior
then the previous algorithms. Even for the relaxed version of the problem,
$S^3$ algorithms are strictly superior to the previous solutions.
In fact, we need at most two samples per window for sequence-based windows and
at most $3\log{n}$ samples for timestamp-based windows plus $4\log^2{n}$ bits.
In contrast, the previous solutions accept memory fluctuations.
Restricting these solutions to the cases where memory bounds are close to ours
creates a bias toward recent elements, resulting in non-uniform sampling.
To illustrate this point, let $p_i$ be a conditional probability of the $N-i$-th element
to be chosen, given that the size of the chain is bounded by $2$.
Then we can show that $p_0/p_n \ge 2$. The bias for priority sampling is smaller;
nevertheless, it is computationally distinguishable for polynomially bounded streams.
Thus, $S^3$ sampling either strictly improves the memory usage or eliminates the non-uniformity of samples (or both).
Both properties are of the great importance for practical applications, as has been pointed out in \cite{networks}.

Second, our results are new in the following sense. As we mentioned
above, deletions introduce additional difficulties: samples expire
 and the size of the domain is unknown. Efficient
algorithms for various streaming models that support deletions exist, such
as dynamic sampling \cite{dynamic}, inverse sampling \cite{inverse}
and biased sampling \cite{biased}. However, all of these results accept a
small probability of failure, either in terms of distribution or in
terms of memory guarantees. $S^3$ algorithms are the \emph{first}
schemas that support $0$ probability of error. We are able to
generate random events with probability $1\over n$, even without
actually knowing the precise value of $n$. Thus, our technique is of
independent interest and can possibly be used in other models that
support deletions.

Probably the most important impact of our results is the
ability to translate any streaming algorithm that is based on
uniform sampling to sliding windows, while preserving worst-case memory
guarantees. To emphasize this point, we present in Section $6$ a
``sample'' of such algorithms. The questions asked there have natural
extensions to sliding windows.
Translation of these algorithms to sliding windows is
straightforward: we replace the underlaying sampling algorithm with
$S^3$. In particular, we address the following problems:
frequency moments, counting triangles in graphs, entropy estimation,
and density estimation. We believe that this is only a small subset of
problems that can be addressed using $S^3$, and thus it may become
a powerful tool in the sliding windows model.

\subsection{A new sampling method -- high-level ideas}
For the sequence-based window, we divide the stream into \emph{buckets}
with a size the same as the window. For each of them we maintain a sample. At
any time, the window intersects up to two buckets, say $B_1$, $B_2$.
For a single sample the algorithm is simple: if the sample from $B_1$
is active, choose it; otherwise choose the sample from $B_2$. Since
the number of expired elements in $B_1$ is equal to the number of
arrived elements in $B_2$, the uniform distribution is preserved. To
create a $k$-sample with replacement, we repeat the procedure $k$
times. We can generalize the idea to a $k$-sample without replacement.
We generate $k$ samples without replacement in every bucket, using the
reservoir algorithm, and combine them as follows. If $i$ samples are
expired in $B_1$, take the $k-i$ active samples from $B_1$ and $i$
samples from $B_2$. Simple analysis shows that the distribution is
uniform.

For the $S^3$ algorithm with replacement from the timestamp-based window, we
maintain a list of buckets, $\zeta$-decomposition. The last bucket,
$B$, may contain both expired and active elements but is smaller
then the union of other buckets. For each bucket we maintain a
$O(1)$-memory structure that contains independent samples from the
bucket and other statistics. We can combine bucket samples with
corresponding probabilities to generate samples of their union.
However, we cannot easily combine last bucket's samples, since the
number of active elements, $n$, is unknown. To overcome this
problem, we exploit the fact that a random sample of $B$ chooses a
(fixed) active element $p$ w.p. ${1\over |B|} \ge {1\over n} $, that
can be reduced to $1\over n$ by generating an independent event w.p.
$|B| \over n$. We prove that it is possible to generate such event
without knowing $n$.

Finally, we show that a $k$-sample without replacement may be
generated from $k$ independent samples, $R_0,\dots, R_{k-1}$, when
$R_i$ samples all but $i$ last active elements. Such samples can be
generated if, in addition, we store last $k$ elements.

Our algorithms generate independent samples for non-overlapping
windows. The independency follows from the nice property of
the reservoir algorithm (that we use to generate samples in the
buckets). Let $R_1$ be a sample generated for the bucket $B$, upon
arrival of $i$ elements of $B$. Let $R_2$ be a fraction of the final
sample (i.e., the sample when the last element of $B$ arrives) that
belongs to the last $|B|-i$ elements. The reservoir algorithm implies
that $R_1$ and $R_2$ are independent. Since the rest of the buckets
contain independent samples as well, we conclude that $S^3$ is
independent for non-overlapping windows.

\subsection{Related work}
Substantial work has been done in the streaming model including
(among many many others) the following papers.
A frequency moments problem was introduced and studied by  Alon, Matias and Szegedy
\cite{ams}, and then by Bhuvanagiri, Ganguly, Kesh and Saha \cite{frequency1},
Bar-Yossef, Jayram, Kumar and Sivakumar
\cite{frequency_lower_bound1}, Chakrabarti, Khot and Sun
\cite{frequency_lower_bound2}, Coppersmith and Kumar
\cite{frequency_impr2}, Ganguly \cite{frequency_impr1}, and
Indyk and Woodruff \cite{frequency}.
Graph algorithms were studied by Bar-Yosseff, Kumar and Sivakumar \cite{triangle2},
Buriol, Frahling, Leonardi, Marchetti-Spaccamela and Sohler \cite{triangle},
Feigenbaum, Kannan, McGregor, Suri and Zhang \cite{distnaces, graphs1}, and
Jowhari and Ghodsi \cite{triangle1}.
Entropy approximation was researched by Chakrabarti, Cormode and McGregor \cite{entropy},
Chakrabarti, Do Ba and Muthukrishnan \cite{entropy1},
Guha, McGregor and Venkatasubramanian \cite{entropy3},
and Lall, Sekar, Ogihara, Xu and Zhang \cite{entropy2}.
Clustering problems were studied by Aggarwal, Han, Wang, and Yu \cite{more-biased},
Guha, Meyerson, Mishra, Motwani and O'Callaghan \cite{clustering}, and Palmer and Faloutsos
\cite{density_sampling}. The problem of estimating the number of distinct elements was addressed by
Bar-Yossef, Jayram, Kumar, Sivakumar and Trevisan \cite{distinct_el},
Cormode, Datar, Indyk, and Muthukrishnan \cite{hamming} and Ganguly \cite{distinct2}.

 Datar, Gionis, Indyk
and Motwani \cite{statistics} pioneered the research in this area,
presenting exponential histograms, effective and simple solutions for
a wide class of functions over sliding windows. In particular, they
gave a memory-optimal algorithm for count, sum, average, $L_p,
p\in[1,2]$ and other functions. Gibbons and Tirthapura
\cite{gibbons} improved the results for sum and count, providing
memory and time-optimal algorithms. Feigenbaum, Kannan and Zhang \cite{diameter} addressed
the problem of computing diameter. Lee and Ting in \cite{counting1}
gave a memory-optimal solution for the relaxed version of the count
problem. Chi, Wang, Yu and Muntz \cite{ucla} addressed a problem of
frequent itemsets. Algorithms for frequency counts and quantiles
were proposed by Arasu and Manku \cite{approx_counters}. Further
improvement for counts was reported by Lee and Ting
\cite{better-freq}. Babcock, Datar, Motwani and O'Callaghan
\cite{variance} provided an effective solution of variance and
$k$-medians problems. Algorithms for rarity and similarity were
proposed by Datar and Muthukrishnan \cite{similarity}.
Golab, DeHaan, Demaine, Lopez-Ortiz and Munro \cite{golab} provided
an effective algorithm for finding frequent elements.
Detailed surveys of recent results can be found in \cite{strbook, strbook1}.

\subsection{Roadmap and notations}
We use the following notations throughout our paper. We denote by $D$ a
stream and by $p_i, i\ge 0$ its $i$-th element. For $0\le x<y$ we
define $[x,y] = \{i, x\le i \le y\}$. Finally, \emph{bucket}
$B(x,y)$ is the set of all stream elements between $p_x$ and
$p_{y-1}$: $B(x,y) = \{p_i,i\in[x,y-1]\}$.

Sections $2$ and $3$ present $S^3$ algorithms for sequence-based
windows, with and without replacement. Section $4$ and
$5$ are devoted to $S^3$ algorithms for timestamp-based windows,
with and without replacement. Section $6$ outlines possible
applications for our approach. Due to the lack of space, some proofs
are omitted from the main body of the paper, but they can all be found in the appendix.

\section{$\textbf{S}^3$ Algorithm With Replacement for Sequence-Based Windows} Let $n$ be the predefined size
of a window. We say that a bucket is \emph{active} if all its
elements have arrived and at least one element is non-expired. We say
that a bucket is \emph{partial} if not all of its elements have arrived. We
show below how to create a single random sample. To create a
$k-$random sample, we repeat the procedure $k$ times, independently.

We divide $D$ into buckets $B(in, (i+1)n), i=0,1,\dots$. At any
point of time, we have exactly one active bucket and at most one
partial bucket. For every such bucket $B$, we independently generate
a single sample, using the reservoir algorithm \cite{reservoir}. We
denote this sample by $X_B$.

Let $B$ be a partial bucket and $C \subseteq B$ be the set of all
arrived elements. The properties of the reservoir algorithm imply that
$X_B$ is a random sample of $C$.

Below, we construct a random sample $Z$ of all non-expired elements.
Let $U$ be the active bucket. If there is no partial bucket, then
$U$ contains only all non-expired elements. Therefore,
$Z=X_U$ is a valid sample. Otherwise, let $V$ be the partial bucket.
Let $ U_e = \{x: x\in U, x \textrm{ is expired}\}, U_a = \{x: x\in
U, x \textrm{ is non-expired}\}, V_a = \{x: x\in V, x \textrm{
arrived} \}.
$

Note that $|V_a| =  |U_e|$ and let $s = |V_a|$. Also, note that our
window is $U_a \cup V_a$ and $X_V$ is a random sample of $V_a$. The
random sample $Z$ is constructed as follows. If $X_U$ is not
expired, we put $Z = X_U$, otherwise $Z=X_V$. To prove the
correctness, let $p$ be a non-expired element. If $p \in U_a$, then
$P(Z=p) = P(X_U = x) = {1\over n}$. If $x \in V_a$, then
$$
P(Z = p) = P(X_U \in U_e, X_V = p) = P(X_U \in U_e)P(X_V = p) =
{s\over n}{1\over s} = {1\over n}.
$$
Therefore, $Z$ is a valid random sample. We need to store only
samples of active or partial buckets. Since the number of such
buckets is at most two and the reservoir algorithm requires $\Theta(1)$
memory, the total memory of our algorithm for $k$-sample is
$\Theta(k)$.

\section{$\textbf{S}^3$ Algorithm Without Replacement for Sequence-Based Windows}
We can generalize the idea above to provide a $k$-random sample
without replacement. In this section $k$-sample means $k$-random
sampling without replacement.

We use the same buckets $B(in, (i+1)n), i=0,1,\dots$. For every such
bucket $B$, we independently generate a $k$-sample $X_B$, using
the reservoir algorithm.

Let $B$ be a partial bucket and $C \subseteq B$ be the set of all
arrived elements. The properties of the reservoir algorithm imply that
either $X_B = C$, if $|C|<k$, or $X_B$ is $k$-sample of $C$. In both
cases, we can generate $i$-sample of $C$ using $X_B$ only, for any
$0 < i \le min(k,|C|)$.

Our algorithm is as follows. Let $U$ be the active bucket. If there
is no partial bucket, then $U$ contains only all active elements.
Therefore, we can put $Z=X_U$. Otherwise, let $V$ be the
partial bucket. We define $ U_e,U_a, V_a, s$ as before and construct
$Z$ as follows. If all elements of $X_U$ are not expired $Z = X_U$.
Otherwise, let $i$ be the number of expired elements, $i = |U_e \cap
X_U|$. As we mentioned before, we can generate an $i$-sample of $V_a$
from $X_V$, since $i\le \min(k, s)$. We denote this sample as
$X_{V}^i$ and put
$$
 Z = (X_U \cap U_a) \cup X_{V}^i.
$$
We will prove now that $Z$ is a valid random sample. Let $Q =
\{p_{j_1}, \dots, p_{j_k} \}$ be a fixed set of $k$ non-expired
elements such that $ j_1 < j_2 <...<j_k$. Let $i = |Q \cap V_A|$, so
$\{p_{j_1},\dots, x_{j_{k-i}}\} \subseteq U_a$ and
$\{p_{j_{k-i+1}},\dots, x_{j_{k}}\} \subseteq V_a$. If $i = 0$, then
$Q \subseteq U$ and
$$
P(Z = Q) = P(X_U = Q) = {1\over {n \choose k}}.
$$
Otherwise, by independency of $X_U$ and $X_{V}^i$
$$
P(Z = Q) = P(|X_U \cap U_e| = i, \{p_{j_{1}},\dots, p_{j_{k-i}}\}
\subseteq X_U, X_{V}^i = \{p_{j_{k-i+1}},\dots, p_{j_k}\} ) =
$$
$$
P(|X_U \cap U_e| = i, \{p_{j_{1}},\dots, p_{j_{k-i}}\}  \subseteq
X_U) P( X_{V}^i = \{p_{j_{k-i+1}},\dots, p_{j_k}\}) = {{s \choose
i}\over {n \choose k}}*{1\over {s \choose i}} ={1\over {n \choose
k}}.
$$

Therefore, $Z$ is a valid random sample of non-expired elements.
Note that we store only samples of active or partial buckets. Since
the number of such buckets is at most two and the reservoir algorithm
requires $O(k)$ memory, the total memory of our algorithm is $O(k)$.

\section{$\textbf{S}^3$ Algorithm With Replacement for Timestamp-Based
Windows}\label{sec:tbwr}

Let $n=n(t)$ be the number of non-expired elements. For each element
$p$, timestamp $T(p)$ represents the moment of $p$'s entrance. For a
window with (predefined) parameter $t_0$, $p$ is active at time $t$
if $t - T(p) < t_0$. We show below how to create a single random
sample. To create a $k-$random sample, we repeat the procedure $k$
times, independently.

\subsection{Notations}
A bucket structure $BS(x,y)$ is a group $\{p_x, x, y, T(x), R_{x,y},
Q_{x,y}, r, q\}$, where $T(x)$ is a timestamp of $p_x$, $R, Q$ are
independent random samples from $B(x,y)$ and $r,q$ are indexes of
the picked (for random samples) elements. We denote by $N(t)$ the
size of $D$ at the moment $t$ and by $l(t)$ the index of the earliest
active element. Note that $N(t) \le N(t+1), l(t) \le l(t+1)$ and
$T(p_i) \le T(p_{i+1})$.

\subsection{$\zeta$-decomposition}

Let $a\le b$ be two indexes. $\zeta$-decomposition of a bucket
$B(a,b)$, $\zeta(a,b)$, is an ordered set of bucket structures with
independent samples inductively defined below.
$$
\zeta(b,b) := BS(b,b+1),
$$
and for $a < b $,
$$
\zeta(a, b) := \left\langle BS(a, c), \zeta(c, b) \right\rangle,
$$
where $ c= a + 2^{\lfloor \log{(b+1-a)}\rfloor - 1}.$

Note that $\left|\zeta(a,b)\right| = O(\log{(b-a)})$, so
$\zeta(a,b)$ uses $O(\log{(b-a)})$ memory.

Given $p_{b+1}$, we inductively define an operator
$Incr(\zeta(a,b))$ as follows.
$$
Incr(\zeta(b,b)):= \left\langle BS(b,b+1), BS(b+1,b+2)\right\rangle.
$$
For $a<b$, we put
$$
Incr(\zeta(a,b)) := \left\langle BS(a,v),
Incr(\zeta(v,b))\right\rangle,
$$
where $v$ is defined below.

If $\lfloor \log(b+2-a) \rfloor = \lfloor \log(b+1-a) \rfloor $ then
we put $v=c$, where $BS(a,c)$ is the first bucket structure of
$\zeta(a,b)$. Otherwise, we put $v = d$, where $BS(c,d)$ is the
second bucket structure of $\zeta(a,b)$. (Note that $\zeta(a,b)$
contains at least two buckets for $a<b$.)

We show how to construct $BS(a,d)$ from $BS(a,c)$ and $BS(c,d)$. We
have in this case $ \lfloor \log(b+2-a) \rfloor = \lfloor
\log(b+1-a) \rfloor +1 $, and therefore $b+1-a = 2^i - 1$ for some
$i\ge 2$. Thus $ c - a = 2^{\left\lfloor
\log{\left(2^i-1\right)}\right\rfloor - 1} = 2^{i-2} $ and
$$
\left\lfloor \log(b+1-c) \right\rfloor = \left\lfloor \log{(b+1-a -
(c-a))} \right\rfloor = \left\lfloor \log
{\left(2^i-2^{i-2}-1\right)} \right\rfloor = i-1.
$$
Thus $ d - c = 2^{\left\lfloor \log (b+1-c) \right\rfloor - 1} =
2^{i-2} =c - a$. Now we can create $BS(a,v)$ by unifying $BS(a,c)$
and $BS(c,d)$: $ BS(a,v) = \left\{p_a, d-a, R_{a,d}, Q_{a,d},
r^\prime, q^\prime\right\}$. We put $R_{a,d} = R_{a,c}$ with probability $1\over 2$ and 
$R_{a,d} = R_{c,d}$ otherwise. Since $d - c = c - a$, and $R_{c,d}, R_{a,c}$ are distributed uniformly, 
we conclude that $R_{a,d}$ is distributed uniformly as well.
$Q_{a,d}$ is defined similarly and $r^\prime,
q^\prime$ are indexes of the chosen samples. Finally, the new
samples are independent of the rest of $\zeta$'s samples. Note also
that $Incr(\zeta(a,b))$ requires $O(\log{(b-a)})$ operations.

\begin{lemma}\label{lm:incr}
For any $a$ and $b$, $ Incr(\zeta(a,b)) = \zeta(a,b+1). $
\end{lemma}

\begin{lemma} \label{lm:2}
For any $t$ with a positive number of active elements, we are able to
maintain one of the following:
\begin{enumerate}
\item{
$
 \zeta(l(t), N(t)),
$ }

 or
\item{
$
 BS(y_t,z_t), \zeta(z_t,N(t)),
$}

where ${y_t} < l(t) \le z_t$, $z_{t} - y_{t} \le N(t) + 1- z_{t}$
and all random samples are independent.
\end{enumerate}

\end{lemma}

\subsection{Sample generation}\label{sub:1}

We use the following notations for this section. Let $B_1= B(a,b)$
and $B_2=B(b,N(t)+1)$ be two buckets such that $p_a$ is expired,
$p_b$ is active and $|B_1| \le |B_2|$. Let $BS_1$ and $BS_2$ be
corresponding bucket structures, with independent random samples
$R_1, Q_1$ and $R_2, Q_2$. We put $\alpha=b-a$ and $\beta =
N(t)+1-b$. Let $\gamma$ be the (unknown) number of non-expired
elements inside $B_1$, so $n=\beta+\gamma$. We stress that $\alpha,
\beta$ are known and $\gamma$ is unknown.

\begin{lemma}\label{lm:zero-one} It is possible to generate a random sample
$Y=Y(Q_1)$ of $B_1$, with the following distribution:
$$
P(Y = p_{b-i}) = {\beta \over (\beta+i)(\beta+i-1)},\ \ \ 0 < i <
\alpha,
$$
$$
P(Y = p_a) = {\beta \over \beta+\alpha-1}.
$$
$Y$ is independent of $R_1,R_2,Q_2$ and can be generated within
constant memory and time, using $Q_1$.
\end{lemma}

\begin{lemma}\label{lm:4}
It is possible to generate a zero-one random variable $X$ such that
$ P(X=1) = {\alpha \over \beta + \gamma}. $ $X$ is independent of
$R_1, R_2, Q_2$ and can be generated using constant time and memory.
\end{lemma}

\begin{lemma}\label{lm:3}
It is possible to construct a random sample $V$ of all non-expired
elements using only the data of $BS_1, BS_2$ and constant time and
memory.
\end{lemma}

\subsection{Main results}

\begin{theorem}
We can maintain a random sample over all non-expired elements using
$\Theta(\log{n})$ memory.
\end{theorem}
\begin{proof} By using lemma \ref{lm:2}, we are able to maintain
one of two cases. If case 1 occurs, we can combine random variables
of all bucket structures with appropriate probabilities and get
a random sample of all non-expired elements. If case 2 occurs, we use
notations of Section \ref{sub:1}, interpret the first bucket as
$B_1$ and combine buckets of $\zeta$-decomposition to generate
samples from $B_2$. Properties of the second case imply $|B_1| \le
|B_2|$ and therefore, by using lemma \ref{lm:3}, we are able to produce a
random sample as well. All procedures, described in the lemmas
require $\Theta(\log{n})$ memory. Therefore, the theorem is correct.
\end{proof}

\begin{lemma}\label{lm:lower_bound}
The memory usage of maintaining a random sample within a timestamp-based
window has a lower bound $\Omega(log(n))$.
\end{lemma}

\section{$\textbf{S}^3$ Algorithm Without Replacement for Timestamp-Based Windows}
Informally, the idea is as follows. We maintain $k$ independent
random samples $R_0,\dots, R_{k-1}$ of active elements, using the
algorithm from Section $4$. The difference between these
samples and the $k$-sample with replacement is that $R_i$ samples
all active elements except the last $i$. This can be done using
$O(k+k\log{n})$ memory. Finally, $k$-sample without replacement can
be generated using $R_1,\dots, R_k$ only.

Let us describe the algorithm in detail. First, we construct
$R_{i}$. To do this, we maintain an auxiliary array with the last $i$
elements. We repeat all procedures in Section \ref{sec:tbwr}, but we
``delay'' the last $i$ elements. An element is added to
$\zeta$-decomposition only when more then $i$ elements arrive after
it. We prove the following variant of Lemma \ref{lm:2}.

\begin{lemma}\label{lm:without}
Let $0<i\le k$. For any $t$ with more then $i$ active elements, we
are able to maintain one of the following:

\begin{enumerate}
\item{
$
 \zeta(l(t), N(t) - i),
$ }

 or
\item{
$
 BS(y_t,z_t), \zeta(z_t,N(t) - i),
$}

where ${y_t} < l(t) \le z_t$ and $z_{t} - y_{t} \le N(t) + 1 - i -
z_{t}$ and all random samples of the bucket structures are
independent.
\end{enumerate}
\end{lemma}
The proof is presented in the appendix. The rest of the procedure
remains the same. Note that we can use the same array for every $i$,
and therefore we can construct $R_0,\dots, R_{k-1}$ using $\Theta(k
+ k\log{n})$ memory.

In the reminder of this section, we show how $R_0,\dots, R_{k-1}$ can be
used to generate a $k$-sample without replacement. We denote by
$R_i^j$ a $i$-random sample without replacement from $[1, j]$.
\begin{lemma}\label{lm:last}
$R_{a+1}^{b+1}$ can be generated using independent $R_a^b$,
$R_1^{b+1}$ samples only.
\end{lemma}

\begin{lemma}\label{lm:lastlast}
$R_{k}^{n}$ can be generated using only independent samples $R_1^n$,
$R_1^{n-1}$,\dots, $R_1^{n-k+1}$.
\end{lemma}
\begin{proof} By using Lemma \ref{lm:last}, we can generate
$R_2^{n-k+2}$ using $R_1^{n-k+1}$ and $R_1^{n-k+2}$. We can repeat
this procedure and generate $R_{j}^{n-k+j}, 2\le j\le k$, using
$R_{j-1}^{n-k+j-1}$ (that we already constructed by induction) and
$R_{1}^{n-k+j}$. For $j=k$ we have $R_k^n$. \end{proof}

By using Lemma \ref{lm:lastlast}, we can generate $k$-sample without replacement
using only $R_0,\dots, R_{k-1}$.

\section{Applications} Consider that algorithm $\Lambda$ is sampling-based,
i.e., it operates on uniformly chosen subset of $D$ instead of the
whole stream. Such an algorithm can be immediately transformed to
sliding windows by replacing the underlying sampling method with
$S^3$. We obtain the following general result and illustrate it with the
examples below.

\begin{corollary}
For the sampling-based algorithm $\Lambda$ that solves problem $P$, there
exists an algorithm $\Lambda'$ that solves $P$ on sliding windows.
The memory guarantees are preserved for sequence-based windows and
have a multiplicative overhead of $\log{n}$ for timestamp-based
windows.
\end{corollary}

Frequency moment is a fundamental problem in data stream processing.
Given a stream of elements, such that $p_j \in [m]$, the frequency
of each $i\in [m]$ is defined as $|j|p_j=i|$ and the $k$-th frequency
moment is defined as $F_k = \sum_{i=1}^m x_i^k$. The first algorithm
for frequency moments for $k>2$ was proposed in the seminal paper of
Alon, Matias and Szegedy \cite{ams}. They present an algorithm that
uses $O(m^{1-{1\over k}})$ memory. Numerous improvements to lower
and upper bounds have been reported, including the works of Bar-Yossef,
Jayram, Kumar and Sivakumar \cite{frequency_lower_bound1},
Chakrabarti, Khot and Sun \cite{frequency_lower_bound2}, Coppersmith
and Kumar \cite{frequency_impr2}, and Ganguly\cite{frequency_impr1}.
Finally, Indyk and Woodruff \cite{frequency} and later Bhuvanagiri,
Ganguly, Kesh and Saha \cite{frequency1} presented algorithms that
use $\tilde{O}(m^{1-{2\over k}})$ memory and are optimal. The
algorithm of Alon, Matias and Szegedy \cite{ams} is sampling-based,
thus we can adapt it to sliding windows using $S^3$. The memory
usage is not optimal, however this is the first algorithm for
frequency moments over sliding windows that works for all $k$.
Recently Braverman and Ostrovsky \cite{our} adapted the algorithm
from \cite{frequency1} to sliding windows, producing a memory-optimal
algorithm that uses $\tilde{O}(m^{1-{2\over k}})$. However, it
involves $k^k$ multiplicative overhead, making it infeasible for
large $k$; thus these results are generally cannot be compared. We have

\begin{corollary}
For any $k>2$, there exists an algorithm that maintains an
approximation of the $k$-th frequency moment over sliding windows using
$\tilde{O}(m^{1-{1\over k}})$ bits.
\end{corollary}

Recently, numerous graph problems were addressed in the streaming
environment. Stream elements represent edges of the
graph, given in arbitrary order. (We refer readers to
\cite{triangle} for a detailed explanation of the model). One of
the fundamental graph problems is estimating a number of small cliques in
a graph, in particular the number of triangles. Effective solutions
were proposed by Jowhari and Ghodsi \cite{triangle1}, Bar-Yosseff,
Kumar and Sivakumar \cite{triangle2} and Buriol, Frahling, Leonardi,
Marchetti-Spaccamela and Sohler \cite{triangle}. The last paper
presented an $(\epsilon, \delta)$-approximation algorithm that uses
$O(1+{\log{|E|}\over |E|}{1\over
\epsilon^2}{|T_1|+2|T_2|+3|T_3|\over |T_3|}\log{2\over \delta})$
memory (\cite{triangle}, Theorem $2$) that is the best result so
far. Here, $|T_i|$ represents the number of node-triplets having $i$
edges in the induced sub-graph. The algorithm is applied on a random
sample collected using the reservoir method. By replacing the reservoir
sampling with $S^3$, we obtain the following result.
\begin{corollary}
There exists an algorithm that maintains an $(\epsilon,
\delta)$-approximation of the number of triangles over sliding
windows. For sequence-based windows it uses $O(1+{\log{|E_W|}\over
|E_W|}{1\over \epsilon^2}{|T_1|+2|T_2|+3|T_3|\over |T_3|}\log{2\over
\delta})$ memory, where $E_W$ is the set of active edges.
Timestamp-based windows adds a multiplicative factor of $\log{n}$.
\end{corollary}

Following \cite{triangle}, our method is also applicable for
incidence streams, where all edges of the same vertex come together.

The entropy of a stream is defined as $H = -\sum_{i=1}^m {x_i\over
N}\log{x_i\over N}$, where $x_i$ is as above. The entropy norm is
defined as $F_H = \sum_{i=1}^m {x_i}\log{x_i}$. Effective solutions
for entropy and entropy norm estimations were recently reported by
Guha, McGregor and Venkatasubramanian \cite{entropy3}, Chakrabarti,
Do Ba and Muthukrishnan \cite{entropy1}, Lall, Sekar, Ogihara, Xu
and Zhang \cite{entropy2} and Chakrabarti, Cormode and McGregor
\cite{entropy}. The last paper presented an algorithm that is based
on a variation of reservoir sampling. The algorithm maintains
entropy using $O(\epsilon^{-2}\log{\delta^{-1}})$ that is nearly
optimal. The authors also considered the sliding window model and
used a variant of priority sampling \cite{sampling} to obtain the
approximation. Thus, the worst-case memory guarantees are not
preserved for sliding windows. By replacing priority sampling with
$S^3$ we obtain

\begin{corollary}
There exists an algorithm that maintains an $(\epsilon,
\delta)$-approximation of entropy on sliding windows using
$O(\epsilon^{-2}\log{\delta^{-1}}\log{n})$ memory.
\end{corollary}

Moreover, $S^3$ can be used with the algorithm from \cite{entropy1}
to obtain $\tilde{O}(1)$ memory for large values of the entropy norm.
This algorithm is based on reservoir sampling and thus can be
straightforwardly implemented in sliding windows. As a result, we
build the first solutions with provable memory guarantees on sliding
windows.

$S^3$ algorithms can be naturally extended to some biased functions.
Biased sampling \cite{biased} is non-uniform, giving larger
probabilities for more recent elements. The distribution is defined
by a biased function. We can apply $S^3$ to implement step biased
functions, maintaining $S^3$ over each window with different
lengths and combining the samples with corresponding probabilities.
Our algorithm can extend the ideas of Feigenbaum, Kannan, Strauss and Viswanathan
\cite{testing}
for testing and spot-checking to sliding windows.
Finally, we can apply $S^3$ to the algorithm of Procopiuc and
Procopiuc for density estimation \cite{density}, since it is based
on the reservoir algorithm as well.

\newpage
 \begin{center}
    {\bf APPENDIX}
  \end{center}
\appendix

\textbf{Lemma \ref{lm:incr}.} \textit{For any $a$ and $b$, $
Incr(\zeta(a,b)) = \zeta(a,b+1). $}
\begin{proof}
We prove the lemma by induction on $b-a$. If $a=b$ then, since $b+1
= b + 2^{\lfloor \log{((b+1)+1-b)}\rfloor - 1}$, we have, by
definition of $\zeta(b, b+1)$,
$$
\zeta(b,b+1) =\left\langle BS(b,b+1), \zeta(b+1,b+1)\right\rangle =
\left\langle BS(b,b+1), BS(b+1,b+2)\right\rangle = Incr(\zeta(b,b)).
$$
We assume that the lemma is correct for $b-a < h$ and prove it for
$b-a = h$. Let $BS(a,v)$ be the first bucket of $Incr(\zeta(a,b))$.
Let $BS(a,c)$ be the first bucket of $\zeta(a,b)$. By definition, if
$\lfloor \log(b+2-a) \rfloor = \lfloor \log(b+1-a) \rfloor $ then
$v=c$. We have
$$
v = c= a + 2^{\lfloor \log{(b+1-a)}\rfloor - 1} = a + 2^{\lfloor
\log{(b+2-a)}\rfloor - 1}.
$$
Otherwise, let $BS(c,d)$ be the second bucket of $\zeta(a,b)$. We
have from above $\lfloor \log(b+2-a) \rfloor = \lfloor \log(b+1-a)
\rfloor +1$, $d-c = c-a$ and $v = d$. Thus
$$
v = d = 2c - a = 2\left(a + 2^{\lfloor
\log{(b+1-a)}\rfloor-1}\right) - a =
 a + 2^{\lfloor \log{(b+1-a)}\rfloor} =
 a + 2^{\lfloor
\log{(b+2-a)}\rfloor - 1}.
$$
In both cases $v = a + 2^{\lfloor \log{((b+1)+1-a)}\rfloor - 1}$ and
, by definition of $\zeta$
$$
\zeta(a, b+1) = \left\langle BS(a,v), \zeta(v, b+1)\right\rangle.
$$
By induction, since $b-v<h$, we have $ Incr\left(\zeta(v,b)\right) =
\zeta(v, b+1). $ Thus
$$
\zeta(a, b+1) = \left\langle BS(a,v), \zeta(v, b+1)\right\rangle =
\left\langle BS(a,v), Incr\left(\zeta(v,b)\right)\right\rangle =
Incr(\zeta(a,b)).
$$
\end{proof}

\textbf{Lemma \ref{lm:2}.} \textit{ For any $t$ with a positive number
of active elements, we are able to maintain one of the following:}
\begin{enumerate}
\item{
\textit{ $ \zeta(l(t), N(t)), $} }

 \textit{ or}
\item{
\textit{ $ BS(y_t,z_t), \zeta(z_t,N(t)), $}}

\textit{ where ${y_t} < l(t) \le z_t$, $z_{t} - y_{t} \le N(t) + 1-
z_{t}$ and all random samples are independent.}
\end{enumerate}

\begin{proof}
We prove the lemma by induction on $t$. First we assume  that $t=0$.
If no element arrives at time $0$, the stream is empty and we do
nothing. Otherwise, we put $\zeta(0,0)= BS(0,1)$, and for any $i,
0<i\le N(0)$ we generate $\zeta(0, i)$ by executing $Incr(\zeta(0,
i-1))$. Therefore, at the end of this step, we have $\zeta(0, N(0))
= \zeta(l(0), N(0))$. So, the case (1) is true.

 We assume that the lemma is correct for $t$ and
prove it for $t+1$.

\begin{enumerate}
\item{

If for $t$ the window is empty, then the procedure is the same as
for the basic case.

}

\item{

If for $t$ we maintain case $(1)$, then we have three sub-cases.
\begin{enumerate}
\item{

If $p_{l(t)}$ is not expired at the moment $t+1$, then $l(t+1) =
l(t)$. Similar to the basic case, we apply $Incr$ procedure for
every new element with index $i, N(t) < i \le N(t+1)$. Due to the
properties of $Incr$, we have at the end $\zeta(l(t+1),N(t+1))$.
Therefore case (1) is true for $t+1$.

}

\item{

 If $p_{N(t)}$ is expired, then our current bucket structures
 represent only expired elements. We delete them and apply the procedure for the basic case.
}

\item{

The last sub-case is the one when $p_{N(t)}$ is not expired and
$p_{l(t)}$ is expired. Let $\langle BS_1,\dots,BS_h\rangle$,
$\left(BS_i = BS(v_i, v_{i+1})\right) $ be all buckets of
$\zeta(l(t), N(t))$. Since $p_{N(t)}$ is not expired, there exists
exactly one bucket structure, $BS_i$, such that $p_{v_i}$ is expired
and $p_{v_{i+1}}$ is not expired. We can find it by checking all the
bucket structures, since we store timestamps for $p_{v_i}$s. We put
$$
y_{t+1} = v_i, z_{t+1} = v_{i+1}.
$$
We have by definition
$$
\zeta(z_{t+1}, N(t)) = \zeta(v_{i+1}, N(t)) = \left\langle
BS_{i+1},\dots,BS_k\right\rangle.
$$
Applying $Incr$ procedure to all new elements, we construct
$\zeta(z_{t+1}, N(t+1))$. Finally, we have:
$$
z_{t+1} - y_{t+1} = v_{i+1} - v_{i} = 2^{\lfloor
\log{(N(t)+1-v_i)}\rfloor - 1}\le
$$
$$
 {1\over 2}(N(t)+1-v_{i}) = {1\over 2}(N(t)+1-y_{t+1}).
$$
Therefore $z_{t+1} - y_{t+1} \le N(t)+1 - z_{t+1} \le N(t+1)+1 -
z_{t+1}$. Thus, case $(2)$ is true for $t+1$. We discard all
non-used bucket structures $BS_1,\dots,BS_{i-1}$. }
\end{enumerate}
}

\item{
 Otherwise, for $t$ we maintain case $(2)$. Similarly, we have three
sub-cases.

\begin{enumerate}
\item{

If $p_{z_t}$ is not expired at the moment $t+1$, we put $y_{t+1} =
y_t, z_{t+1} = z_t$. We have
$$
 z_{t+1} - y_{t+1} = z_{t} - y_{t} \le
N(t)+1-z_{t} \le N(t+1)+1 - z_{t+1}.
$$
Again, we add the new elements using $Incr$ procedure and we
construct $\zeta(z_{t+1}, N(t+1))$. Therefore case $(2)$ is true for
$t+1$.
 }

\item{
 If $p_{N(t)}$ is expired, we apply exactly the same procedure as for
 $2.b$.
 }
\item{
 If $p_{z_t}$ is expired and $p_{N(t)}$ is not expired, we apply exactly the same procedure as for
 $2.c$.
 }
\end{enumerate}
}
\end{enumerate}
 Therefore, the lemma is correct.
\end{proof}

\textbf{Lemma \ref{lm:zero-one}.} \textit{It is possible to generate
a random sample $Y=Y(Q_1)$ of $B_1$, with the following
distribution:
$$
P(Y = p_{b-i}) = {\beta \over (\beta+i)(\beta+i-1)},\ \ \ 0 < i <
\alpha,
$$
$$
P(Y = p_a) = {\beta \over \beta+\alpha-1}.
$$
$Y$ is independent of $R_1,R_2,Q_2$ and can be generated within
constant memory and time, using $Q_1$. }

\begin{proof}
Let $\{H_j\}_{j=1}^{\alpha-1}$ be  a set of zero-one independent
random variables such that
$$
P(H_j = 1) =
{\alpha\beta \over (\beta+j)(\beta+j-1)}.
$$
Let $D = B_1\times \{0,1\}^{\alpha-1}$ and $Z$ be the random vector
with values from $D$, $ Z = \langle Q_1, H_{1},...,H_{\alpha - 1}
\rangle. $ Let $\{A_i\}_{i=1}^{\alpha}$ be a set of subsets of $D$:
$$
A_i = \{\langle q_{b-i}, a_1,\dots, a_{i-1}, 1, a_{i+1},\dots,a_{\alpha - 1}\rangle \ | \ \ a_j \in \{0,1\},j\neq i\}.
$$
Finally we define $Y$ as follows
$$
Y =\left\{\begin{array}{ll} q_{b-i},  &\textrm{if\ } Z \in A_i, \ \ 1\le i < \alpha,\\
                    q_a, & \textrm{\ otherwise}.\\
    \end{array}\right.
$$
Since $Q_1$ is independent of $R_1, R_2, Q_2$, $Y$ is independent of
them as well. We have
$$
P(Y=p_{b-i}) = P(Z \in A_i) = P(Q_1 = q_{b-i}, H_i = 1, H_j \in
\{0,1\} \textrm{\ for\ } j\neq i)=
$$
$$
P(Q_1 = q_{b-i})P(H_i = 1)P( H_j \in \{0,1\}  \textrm{\ for\ } j\neq
i)=
$$
$$
P(Q_1 = q_{b-i})P(H_i = 1)=
 {1\over \alpha}{\alpha\beta \over
(\beta+i)(\beta+i-1)} = {\beta \over (\beta+i)(\beta+i-1)}.
$$
Also,
$$
P(Y=p_a) = 1 - \sum_{i=1}^{\alpha- 1} P(Y=p_{b-i}) = 1 -
\sum_{i=1}^{\alpha- 1} {\beta \over (\beta+i)(\beta+i-1)} =
$$
$$
1 - \beta\sum_{i=1}^{\alpha - 1} \left({1 \over \beta+i-1} - {1
\over \beta+i}\right) = 1 - \beta \left({1 \over \beta} - {1 \over
\beta+\alpha-1}\right) = {\beta \over \beta+\alpha-1}.
$$
By definition of $A_i$, the value of $Y$ is uniquely defined by
$Q_1$ and exactly one $H$. Therefore, the generation of the whole
vector $Z$ is not necessary. Instead, we can calculate $Y$ by the
following simple procedure. Once we know the index of $Q_1$'s value,
we generate the corresponding $H_i$ and calculate the value of $Y$.
We can omit the generation of other $H$s, and therefore we need
constant time and memory.
\end{proof}

\textbf{Lemma \ref{lm:4}.} \textit{ It is possible to generate a zero-one random variable $X$ such that
$ P(X=1) = {\alpha \over \beta + \gamma}. $ $X$ is independent of
$R_1, R_2, Q_2$ and can be generated using constant time and memory.
}
\begin{proof}
Since $\gamma$ is unknown, it cannot be generated by flipping a coin;
a slightly more complicated procedure is required.

Let $Y(Q_1)$ be the random variable from Lemma \ref{lm:zero-one}. We
have
$$
P(\textrm{Y is not expired}) = \sum_{i=1}^\gamma {P(Y = q_{b-i})} =
\sum_{i=1}^\gamma {\beta \over (\beta+i)(\beta+i-1)} =
$$
$$
\beta\sum_{i=1}^\gamma \left({1\over \beta+i-1}- {1\over
\beta+i}\right) = \beta\left({1 \over \beta} - {1 \over \beta +
\gamma}\right) = {\gamma \over \beta + \gamma}.
$$
Therefore $P(\textrm{Y is expired}) = {\beta \over \beta + \gamma}$.

Let $S$ be a zero-one variable, independent of $R_1, R_2, Q_2, Y$ such
that
$$
P(S =1)={\alpha\over \beta}.
$$
We put
$$
X =\left\{\begin{array}{ll} 1,  &\textrm{if\ } Y \textrm{is\ expired\ AND\ } S = 1,\\
                    0, & \textrm{\ otherwise}.\\
    \end{array}\right.
$$
We have
$$
P(X=1) = P(Y \textrm{\ is expired}, S=1) = P(Y \textrm{\ is
expired})P(S=1)= {\beta \over \beta + \gamma}{\alpha\over \beta}=
{\alpha \over \beta + \gamma}.
$$

Since $Y$ and $S$ are independent of $R_1, R_2, Q_2$, $X$ is
independent of them as well. Since we can determine if $Y$ is expired
within constant time, we need a constant amount of time and memory.
\end{proof}

\textbf{Lemma \ref{lm:3}.} \textit{ It is possible to construct a random sample $V$ of all non-expired
elements using only the data of $BS_1, BS_2$ and constant time and
memory.}
\begin{proof} Our goal is to generate a random variable $V$
that chooses a non-expired element w.p. ${1\over \beta+\gamma}$. Let
$X$ be the random variable generated in the previous lemma. We define
$V$ as follows.
$$
V =\left\{\begin{array}{ll} R_1, & R_1 \textrm{\ is not  expired AND\ }  X = 1,\\
                    R_2, & \textrm{otherwise}.\\
    \end{array}\right.
$$

Let $p$ be a non-expired element.
If $p \in B_1$, then since $X$ is independent of $R_1$, we have
$$
P(V=p) = P(R_1 = p, X=1) = P(R_1=p)P(X=1) ={1\over
\alpha}{\alpha\over \beta + \gamma}={1\over \beta + \gamma}= {1\over
n}.
$$
If $p \in B_2$, then
$$
 P(V=p) = (1 - P(R_1 \textrm{\ is not
expired})P(X=1))P(R_2=p)=  \left(1-{\gamma \over \alpha}{\alpha
\over \beta + \gamma}\right){1\over \beta} ={1\over \beta + \gamma}
= {1\over n}.
$$

\end{proof}

\textbf{Lemma \ref{lm:lower_bound}.} \textit{ The memory usage of
maintaining a random sample within the time-based window has a lower bound
$\Omega(log(n))$.}
\begin{proof} Let $D$ be a stream with the following
property. For timestamp $i, 0\le i\le 2t_0$, we have $2^{2t_0-i}$
elements and for $i>2t_0$, we have exactly one element per
timestamp.

For timestamp $0\le i  \le t_0$, the probability to choose $p$ with
$T(p) = i$ at the moment $t_0+i-1$ is
$$
{2^{2t_0-i} \over \sum_{j=i}^{i+t_0-1} 2^{2t_0-j}} = {2^{2t_0-i}
\over
 2^{t_0-i+1} \sum_{j=0}^{t_0-1} 2^{t_0-j-1}}={2^{t_0-1}\over \sum_{j=0}^{t_0-1}
 2^{j}} ={2^{t_0-1}\over 2^{t_0} -1 } > {1\over 2}.
$$
Therefore, the expected number of distinct timestamps that will be
picked between moments $t_0-1$ and $2t_0-1$ is at least $
\sum_{i=t_0-1}^{2t_0-1} {1\over 2} = {t_0+1\over 2}. $ So, with a
positive probability we need to keep in memory at least ${t_0\over
2}$ distinct elements at the moment $t_0$. The number of active
elements $n$ at this moment is at least $2^{t_0}$. Therefore the
memory usage at this moment is $\Omega(\log{n})$, with positive
probability. We can conclude that $\log(n)$ is a lower bound for
memory usage.
\end{proof}

\textbf{Lemma \ref{lm:without}.} \textit{ Let $0<i\le k$. For any
$t$ with more then $i$ active elements, we are able to maintain one
of the following:}

\begin{enumerate}
\item{
\textit{ $ \zeta(l(t), N(t) - i), $}}

 \textit{ or}
\item{\textit{
$
 BS(y_t,z_t), \zeta(z_t,N(t) - i),
$}}

\textit{ where ${y_t} < l(t) \le z_t$ and $z_{t} - y_{t} \le N(t) +
1 - i - z_{t}$ and all random samples of the bucket structures are
independent.}
\end{enumerate}

\begin{proof}
 The proof is the same
as in lemma $4.2$, except for cases $1, 2.b, 3.b$. For these cases, when
the current window is empty, we keep it empty unless more then $i$
elements are active. We can do this using our auxiliary array. Also,
when new elements arrive, some of them may be expired already (if we
kept them in the array). We therefore cannot apply $Incr$ procedure
for any ``new'' element. Instead, we should first skip all expired
elements and then apply $Incr$. The rest of the proof remains the
same.
\end{proof}

\textbf{Lemma \ref{lm:last}.} \textit{ $R_{a+1}^{b+1}$ can be
generated using independent $R_a^b$, $R_1^{b+1}$ samples only.}
\begin{proof} The algorithm is as follows.
$$
R_{a+1}^{b+1} =\left\{\begin{array}{ll}
                    R_a^b \cup \{b+1\}, & \textrm{if } R_1^{b+1} \in R_a^{b}, \\
                    R_a^b \cup R_1^{b+1}, &\rm{otherwise\ } .\\
    \end{array}\right.
$$

Let $X = \{x_1, \dots, x_{a+1}\}$ be a set of points from $[1,b+1]$,
such that $x_1 < x_2 < \dots < x_a < x_{a+1}$.

If $x_{a+1} < b+1$, then we have
$$
P(R_{a+1}^{b+1}=X) = P\left(\bigcup_{j=1}^{a+1} (R_1^{b+1} = x_j
\cap R_a^b = X\backslash \{x_j\})  \right) =
$$
$$
\sum_{j=1}^{a+1}P(R_1^{b+1} = x_j)P(R_a^b = X\backslash \{x_j\})  =
(a+1){1\over b+1}{1\over{b \choose a}} = {1\over{b+1 \choose a+1}}.
$$

Otherwise,
$$
P(R_{a+1}^{b+1}=X) = P\left(R_a^b = X\backslash \{b+1\}, R_1^{b+1}
\in X \right) = {1\over{b \choose a}}{a+1\over b+1} ={1\over{b+1
\choose a+1}}.
$$
\end{proof}


\begin{thebibliography}{99}
\footnotesize
\bibitem{strbook1}
C. Aggarwal (editor), \emph{Data Streams: Models and Algorithms},
\emph{Springer Verlag}, 2007.

\bibitem{biased}
C. Aggarwal, ``On biased reservoir sampling in the presence of stream
evolution'', \emph{Proceedings of the 32nd international conference
on Very large data bases}, pp. 607--618, 2006.

\bibitem{more-biased}
C. Aggarwal, J. Han, J. Wang, P. Yu, ``A Framework for High
Dimensional Projected Clustering of Data Streams''. \emph{VLDB
Conference Proceedings}, pp. 852--863, 2004.

\bibitem{priority1}
N. Alon, N. Duffield, C. Lund, M. Thorup, ``Estimating arbitrary
subset sums with few probes''. \emph{Proceedings of the twenty-fourth
ACM SIGMOD-SIGACT-SIGART symposium on Principles of database
systems}, pp. 317--325, 2005.

\bibitem{ams}
N. Alon, Y. Matias, M.Szegedy, ``The space complexity of
approximating the frequency moments''. \emph{Proceedings of the
twenty-eighth annual ACM symposium on Theory of computing}, pp.
20--29, 1996.

\bibitem{approx_counters} A. Arasu, G. S. Manku,
``Approximate counts and quantiles over sliding windows'',
\emph{Proceedings of the twenty-third ACM SIGMOD-SIGACT-SIGART
symposium on Principles of database systems}, 2004.

\bibitem{models_issues} B. Babcock, S. Babu, M. Datar, R. Motwani, J. Widom,
``Models and issues in data stream systems'', \emph{Proceedings of the
twenty-first ACM SIGMOD-SIGACT-SIGART symposium on Principles of
database systems}, 2002.

\bibitem{sampling} B. Babcock, M. Datar, R. Motwani, ``Sampling from a moving window over streaming data'',
\emph{Proceedings of the thirteenth annual ACM-SIAM symposium on
Discrete algorithms}, pp.633--634, 2002.

\bibitem{variance}B. Babcock, M. Datar, R. Motwani, L. O'Callaghan,
``Maintaining variance and k-medians over data stream windows'',
\emph{Proceedings of the twenty-second ACM SIGMOD-SIGACT-SIGART
symposium on Principles of database systems}, pp.234--243, 2003.

\bibitem{frequency_lower_bound1}
Z. Bar-Yossef, T. S. Jayram, R. Kumar, D. Sivakumar, ``An Information
Statistics Approach to Data Stream and Communication Complexity'',
\emph{Proceedings of the 43rd Symposium on Foundations of Computer
Science}, pp. 209--218, 2002.

\bibitem{triangle2}
Z. Bar-Yosseff, R. Kumar, D. Sivakumar, ``Reductions in streaming
algorithms, with an application to counting triangles in graphs'',
\emph{Proceedings of the thirteenth annual ACM-SIAM symposium on
Discrete algorithms}, pp.623--632, 2002.

\bibitem{distinct_el}
Z. Bar-Yossef, T. S. Jayram, R. Kumar, D. Sivakumar, L. Trevisan,
``Counting Distinct Elements in a Data Stream'',
\emph{Proceedings of the 6th International Workshop on Randomization and Approximation Techniques}, pp.1-10, 2002.

\bibitem{frequency1}
L. Bhuvanagiri, S. Ganguly, D. Kesh, C. Saha, ``Simpler algorithm
for estimating frequency moments of data streams'', \emph{Proceedings
of the seventeenth annual ACM-SIAM symposium on Discrete algorithm},
pp.708--713, 2006.

\bibitem{our}
V. Braverman, R. Ostrovsky, ``Smooth histograms on stream windows'',
\emph{Proceedings of the 48th Symposium on Foundations of Computer
Science}, 2007.

\bibitem{triangle}L. S. Buriol, G. Frahling, S. Leonardi, A. Marchetti-Spaccamela, C. Sohler,
``Counting triangles in data streams'', \emph{Proceedings of the
twenty-fifth ACM SIGMOD-SIGACT-SIGART symposium on Principles of
database systems}, pp.253--262, 2006.

\bibitem{entropy}
A. Chakrabarti, G. Cormode, A. McGregor, ``A near-optimal algorithm
for computing the entropy of a stream''. \emph{In Proceedings of
ACM-SIAM Symposium on Discrete Algorithms}, 2007.

\bibitem{entropy1}
A. Chakrabarti, K. Do Ba, S. Muthukrishnan, ``Estimating Entropy and
Entropy Norm on Data Streams'',  \emph{In Proceedings of the 23rd
International Symposium on Theoretical Aspects of Computer Science},
2006.

\bibitem{frequency_lower_bound2}
A. Chakrabarti, S. Khot, X. Sun, ``Near-optimal lower bounds on the
multi-party communication complexity of set-disjointness'',
\emph{Proceedings of the 18th Annual IEEE Conference on
Computational Complexity}, 2003.


\bibitem{joins}   S. Chaudhuri, R. Motwani, V. Narasayya,
``On random sampling over joins'', \emph{Proceedings of the 1999 ACM
SIGMOD international conference on Management of data}, pp.263-274,
1999.

\bibitem{ucla} Y. Chi, H. Wang, P. S. Yu, R. R. Muntz, ``Moment:
Maintaining Closed Frequent Itemsets over a Stream Sliding Window'',
  \emph{Fourth IEEE International Conference on Data
Mining} (ICDM'04), pp. 59--66, 2004.

\bibitem{hamming} G. Cormode, M. Datar, P. Indyk, S. Muthukrishnan,
``Comparing Data Streams Using Hamming Norms (How to Zero In)'',
\emph{ IEEE Transactions on Knowledge and Data Engineering}, v.15
n.3, pp.529--540,  2003.

\bibitem{inverse}
 G. Cormode, S. Muthukrishnan, I. Rozenbaum,
 ``Summarizing and mining inverse distributions on data streams via dynamic inverse sampling'',
 \emph{Proceedings of the 31st international conference on Very large data bases},
 2005.

\bibitem{frequency_impr2}
D. Coppersmith, R. Kumar, ``An improved data stream algorithm for
frequency moments'', \emph{Proceedings of the fifteenth annual
ACM-SIAM symposium on Discrete algorithms}, pp.151--156, 2004.

\bibitem{statistics} M. Datar, A. Gionis, P. Indyk, R. Motwani,
``Maintaining stream statistics over sliding windows: (extended
abstract)'', \emph{Proceedings of the thirteenth annual ACM-SIAM
symposium on Discrete algorithms}, pp.635--644, 2002.

\bibitem{similarity}M. Datar, S. Muthukrishnan, ``Estimating Rarity and Similarity over Data Stream Windows'',
\emph{Proceedings of the 10th Annual European Symposium on
Algorithms},
 pp.323--334, 2002.

\bibitem{sampling_networks}
N. Duffield, ``Sampling for passive internet measurement: a review'', \emph{Statistical Science}, 19(3), 2004.

\bibitem{sampling_networks2}
N. Duffield, C. Lund, M. Thorup, "Flow sampling under hard resource constraints",
\emph{ACM SIGMETRICS Performance Evaluation Review}, v.32 n.1, 2004.

\bibitem{priority2} N. Duffield, C. Lund and M. Thorup,
``Sampling to estimate arbitrary subset sums'',
\emph{http://www.citebase.org/abstract?id=oai:arXiv.org:cs/0509026},
2005.

\bibitem{diameter}
J. Feigenbaum, S. Kannan, and J. Zhang,
``Computing diameter in the streaming and sliding-window models'',
\emph{Algorithmica}, 41:25--41, 2005.

\bibitem{distnaces}
J. Feigenbaum, S. Kannan, A. McGregor, S. Suri, J. Zhang,
``Graph distances in the streaming model: the value of space'', \emph{SODA}, 2005.

\bibitem{graphs1}
J. Feigenbaum, S. Kannan, A. McGregor, S. Suri, J. Zhang,
``On graph problems in a semi-streaming model'', \emph{Theor. Comput. Sci.}, 2005.

\bibitem{testing}
J. Feigenbaum, S. Kannan, M. Strauss, M. Viswanathan,
``Testing and Spot-Checking of Data Streams'', \emph{Algorithmica}, 34(1): 67-80, 2002.

\bibitem{dynamic}
G. Frahling, P. Indyk, C. Sohler, ``Sampling in dynamic data
streams and applications'', \emph{Proceedings of the twenty-first
annual symposium on Computational geometry}, 2005.

\bibitem{frequency_impr1}
S. Ganguly. ``Estimating Frequency Moments of Update Streams using
Random Linear Combinations''. \emph{Proceedings of the 8th
International Workshop on Randomized Algorithms}, pp. 369-–380, 2004.

\bibitem{distinct2}
S. Ganguly, ``Counting distinct items over update streams'', \emph{Theoretical Computer Science}, pp.211--222, 2007.

\bibitem{mg}  P. B. Gibbons, Y. Matias,
``New sampling-based summary statistics for improving approximate query answers'',
\emph{Proceedings of the 1998 ACM SIGMOD international conference on
Management of data}, pp.331--342, 1998.


\bibitem{gibbons}
P. B. Gibbons, S. Tirthapura, ``Distributed streams algorithms for
sliding windows'', \emph{Proceedings of the fourteenth annual ACM
symposium on Parallel algorithms and architectures}, pp.10--13, 2002.


\bibitem{golab}
L. Golab, D. DeHaan, E. D. Demaine, A. Lopez-Ortiz, J. I. Munro,
"Identifying frequent items in sliding windows over on-line packet streams",
\emph{Proceedings of the 3rd ACM SIGCOMM conference on Internet measurement}, 2003.

\bibitem{entropy3}
S. Guha, A. McGregor, S. Venkatasubramanian, ``Streaming and
sublinear approximation of entropy and information distances'',
\emph{Proceedings of the seventeenth annual ACM-SIAM symposium on
Discrete algorithm}, pp.733-742, 2006.

\bibitem{clustering}
S. Guha, A. Meyerson, N. Mishra, R. Motwani, L. O'Callaghan,
``Clustering Data Streams: Theory and Practice'', \emph{IEEE Trans. on
Knowledge and Data Engineering}, vol. 15, 2003.

\bibitem{frequency}
P. Indyk, D. Woodruff, ``Optimal approximations of the frequency
moments of data streams'', \emph{Proceedings of the thirty-seventh
annual ACM symposium on Theory of computing}, pp.202--208, 2005.

\bibitem{reservoir2}
M. Kolonko, D. W\"{a}sch, ``Sequential reservoir sampling with a
nonuniform distribution'', v.32, i.2, pp.257--273, 2006.

\bibitem{triangle1}
H. Jowhari, M. Ghodsi, ``New streaming algorithms for counting
triangles in graphs'', \emph{Proceedings of the 11th COCOON}, pp.
710--716, 2005.

\bibitem{entropy2}
A. Lall, V. Sekar, M. Ogihara, J. Xu, H. Zhang, ``Data streaming
algorithms for estimating entropy of network traffic'', \emph{ACM
SIGMETRICS Performance Evaluation Review}, pp. 145--156, 2006.

\bibitem{better-freq}
 L. K. Lee, H. F. Ting,
``Frequency counting and aggregation: A simpler and more efficient
deterministic scheme for finding frequent items over sliding
windows'', \emph{ Proceedings of the twenty-fifth ACM
SIGMOD-SIGACT-SIGART symposium on Principles of database systems}
(PODS '06), pp. 290--297, 2006.

\bibitem{counting1} L. K. Lee, H. F. Ting, ``Maintaining significant stream statistics over sliding windows'',
\emph{Proceedings of the seventeenth annual ACM-SIAM symposium on
Discrete algorithm}, pp.724--732, 2006.

\bibitem{reservoir1}
K. Li, ``Reservoir-sampling algorithms of time complexity $O(n(1 +
log(N/n)))$'', \emph{ACM Transactions on Mathematical Software
(TOMS)}, v.20 n.4, pp.481--493, Dec. 1994.

\bibitem{approx_freq} G. S. Manku, R. Motwani, ``Approximate frequency counts over data streams''.
\emph{In Proceedings of the 28th International Conference on Very
Large Data Bases}, pp.356--357, 2002.

\bibitem{strbook} S. Muthukrishnan, ``Data Streams: Algorithms And Applications'' \emph{Foundations and Trends in
Theoretical Computer Science}, Volume 1, Issue 2.

\bibitem{density_sampling} C. R. Palmer, C. Faloutsos,
``Density biased sampling: an improved method for data mining and
clustering'', \emph{Proceedings of the 2000 ACM SIGMOD international
conference on Management of data},
 pp.82--92, 2000

\bibitem{networks}
V. Paxson, G. Almes, J. Mahdavi, M. Mathis, "Framework for IP performance metrics", RFC 2330, 1998.

\bibitem{density}
C. Procopiuc, O. Procopiuc, ``Density Estimation for Spatial Data
Streams'', \emph{Proceedings of the 9th International Symposium on
Spatial and Temporal Databases }, pp.109--126, 2005.

\bibitem{dlt} M. Szegedy, ``The DLT priority sampling is essentially optimal'',
\emph{Proceedings of the thirty-eighth annual ACM symposium on
Theory of computing}, pp.150--158, 2006.

\bibitem{reservoir} J. S. Vitter, ``Random sampling with a reservoir'',
\emph{ACM Transactions on Mathematical Software} (TOMS),
 v.11 n.1, pp.37--57, 1985.


\bibitem{sampling1}
L. Zhang, Z. Li, M. Yu, Y. Wang, Y. Jiang, "Random sampling algorithms for sliding windows over data streams",
\emph{Proc. of the 11th Joint International Computer Conference}, pp. 572--575, 2005.

\end{thebibliography}
\end{document}